\documentclass[article]{elsart}
\usepackage{amssymb}
\usepackage{epsfig}
\begin{document}

\begin{frontmatter}

\title{ Measurement of the Transverse-Longitudinal Cross Sections
in the $p (\vec{e},e^{\prime}p)\pi^{0}$ Reaction in the
$\Delta$ Region}

\author[bmit]{C.~Kunz},
\author[iasa]{N.I.~Kaloskamis},
\author[bmit]{M.O.~Distler\thanksref{distler}},
	\thanks[distler]{Now at Universit\"at, Mainz, Germany.}
\author[bmit]{Z.-L.~Zhou},
\author[asu]{R.~Alarcon},
\author[bmit]{D.~Barkhuff},
\author[bmit]{A.~M.~Bernstein\corauthref{cor}},
        \corauth[cor]{Corresponding author.}
        \ead{bernstein@lns.mit.edu}
\author[bmit]{W.~Bertozzi},
\author[unh]{J.~Calarco},
\author[bmit]{F.~Casagrande},
\author[bmit]{J.~Chen},
\author[asu]{J.~Comfort},
\author[bmit]{G.~Dodson\thanksref{gdodson}},
	\thanks[gdodson]{Now at ORNL, Oak Ridge, TN 37831.}
\author[fsu]{A.~Dooley},
\author[bmit]{K.~Dow},
\author[bmit]{M.~Farkhondeh},
\author[iasa]{S.~ Georgakopoulos},
\author[bmit]{S.~Gilad},
\author[umass]{R.~Hicks},
\author[umass]{A.~Hotta},
\author[umass]{X.~Jiang\thanksref{xdjiang}},
	\thanks[xdjiang]{Now at Rutgers University, Piscataway, NJ 08855.}
\author[iasa]{A.~Karabarbounis},
\author[bmit]{S.~Kowalski},
\author[csu]{D.J.~Margaziotis},
\author[asu,iasa]{C.~Mertz},
\author[umass]{R.~Miskimen},
\author[bmit]{I.~Nakagawa},
\author[iasa]{C.N.~Papanicolas},
\author[bmit]{M.M.~Pavan\thanksref{mpavan}},
	\thanks[mpavan]{Now at TRIUMF, Vancouver, Canada.}
\author[umass]{G.~Peterson},
\author[asu]{A.~Ramirez},
\author[bmit]{D.~Rowntree},
\author[fsu]{A.J.~Sarty\thanksref{as}},
	\thanks[as]{Now at St. Mary's Univ., Halifax, Canada, B3H 3C3.}
\author[umass]{J.~Shaw},
\author[asu]{E.~Six},
\author[iasa]{N.~Sparveris},
\author[bmit]{S.-B.~Soong},
\author[iasa]{S.~Stiliaris},
\author[tu]{T.~Tamae},
\author[bmit]{D.~Tieger},
\author[bmit]{C.~Tschalaer},
\author[bmit]{G.~Tsentalovich},
\author[bmit]{W.~Turchinetz},
\author[iasa]{C.E.~Vellidis},
\author[bmit]{G.A.~Warren\thanksref{glen}},
	\thanks[glen]{Now at Universitat Basel, Switzerland.}
\author[uiuc]{S.~Williamson},
\author[asu]{A.~Young},
\author[bmit]{J.~Zhao},
and
\author[bmit]{T.~Zwart}
\author[]{(The MIT-Bates OOPS Collaboration)}

\address[bmit]{Physics Department and Laboratory for Nuclear
	Science, Massachusetts Institute of Technology, Cambridge,
	MA 02139, USA} 
\address[iasa]{Institute of Accelerating Systems and Applications
	and Department of Physics, University of Athens, Athens,
	Greece} 
\address[asu]{Department of Physics and Astronomy,
        Arizona State University, Tempe, AZ 85287, USA} 
\address[umass]{Department of Physics,
         University of Massachusetts at Amherst, Amherst, MA
	01003, USA} 
\address[fsu]{Department of Physics,
        Florida State University, Tallahassee, FL 32306, USA} 
\address[csu]{Department of Physics and Astronomy,
        California State University at Los Angeles,
        Los Angeles, CA 90032, USA} 
\address[tu]{Laboratory of Nuclear Science,
            Tohoku University, Sendai 982-0826, Japan}
\address[uiuc]{Nuclear Physics Laboratory,
        University of Illinois, Urbana, IL 61820, USA}
\address[unh]{Department of Physics, University of New Hampshire,
	Durham, New Hampshire 03824, USA} 

\date{\today}

\begin{abstract} 
Accurate measurements of the $p(\vec{e},e^{\prime}p)\pi^{0}$ reaction were performed at
$Q^2=0.127$ (GeV/c)$^2$ in the $\Delta$ resonance energy region. The
experiments at the MIT-Bates
Linear Accelerator used an 820 MeV polarized electron beam with the out of plane magnetic
spectrometer system (OOPS). In this paper we report the first simultaneous
determination of both the TL and  TL$^{\prime}$ (``fifth" or polarized)
cross sections at low $Q^{2}$ where the pion cloud contribution is predicted to dominate the quadrupole amplitudes (E2 and C2). The real and imaginary parts of the transverse-longitudinal cross section provide both a
sensitive determination of the Coulomb quadrupole amplitude and a test of
reaction calculations. Comparisons with model calculations are presented. The empirical MAID calculation gives the best overall agreement with this accurate data. The parameters of this model for the values of the resonant multipoles are $|M_{1+}(I=3/2)|= (40.9 \pm 0.3)10^{-3}/m_{\pi}, CMR= C2/M1= -6.5 \pm 0.3\%, EMR=E2/M1=-2.2 \pm 0.9 \%$, where the errors are due to the experimental uncertainties. 

\end{abstract}

\end{frontmatter}


Experimental confirmation of the deviation of the nucleon shape from
spherical symmetry is of fundamental significance and has been the subject
of intense investigation~\cite{Nstar}
since this possibility was originally raised by
Glashow~\cite{Glashow}. For the $J = 1/2$ nucleon, this
determination has focused on the determination of the electric
and Coulomb quadrupole amplitudes (E2, C2) in the
predominantly M1(magnetic dipole-quark spin flip)
$\gamma^{*} N \rightarrow \Delta$ transition. Thus measurements of the E2 and C2 amplitudes represent deviations from spherical symmetry of the $N,\Delta$ system and not the nucleon alone. The experimental difficulty
is that the  E2/M1 and C2/M1 ratios are small (typically $\simeq$ $-2$ to
$-8$ \% at low four momentum transfered, $Q^{2}$). In this case the 
non-resonant (background) and resonant quadrupole amplitudes are the 
same order of
magnitude. Therefore experiments have to be designed to attain the
required precision to separate the signal and background contributions.
This has been accomplished for photo-pion reactions using polarized photon
beams~\cite{LEGS,Mainz}. 

Observation of the deviation from spherical symmetry in pion
electroproduction is more pronounced than in photoproduction.
This is due to the interference between the longitudinal C2 and the
dominant M1 amplitudes in the $\sigma_{TL}$ cross section
\cite{Mertz}. On the other hand the presence of the additional
longitudinal multipoles means that there are more multipoles to
determine and therefore more extensive data must be taken.  The
experiments for an extensive data-base that would allow a model
independent analysis have just begun~\cite{Nstar,CNP,AB,CEBAF}.
At the present time one must rely on reaction
models to extract the resonant  M1, E2, and C2 amplitudes of interest from
the data. As has been pointed out in previous publications, the model
error can be much larger than the experimental error~\cite{Mertz,CNP,Vellidis}. 
Therefore it is important to test model calculations for a
range of center of mass(CM) energies $W$ in the region of 1232 MeV, the
$\Delta$ resonant energy, which provides a range of
background and resonant amplitudes. It is also important to determine specific
cross sections (e.g. $\sigma_{\rm TL}, \sigma_{\rm TL^{'}}$) which are
primarily sensitive to the C2 and background amplitudes respectively. 
In this letter we present the first measurement that
provides such information at and below the resonance energy at low
$Q^{2}$. This requires a polarized electron beam and out
of plane hadron detection. 

In the constituent quark model the d state admixtures in the nucleon and $\Delta$ wave
functions are caused  by the hyperfine
tensor interaction between quarks~\cite{Isgur}. The 
pion cloud contribution to the nucleon and $\Delta$ structure is primarily in the p
wave. This is due to the spontaneously broken chiral
symmetry of QCD in which the pion, an almost Goldstone
Boson, interacts with hadrons via gradient coupling~\cite{Goldstone}.
Therefore it is not surprising~\cite{AB,SL,KY} that model calculations have shown that at
low $Q^{2}$ the pion cloud
contributes significantly to the M1 amplitude and dominates
the E2 and C2 contributions to the
$\gamma^{*} N \rightarrow \Delta$ transition.
The present experiment was performed at  $Q^{2} = 0.127$
(GeV/c)$^{2}$ which is close to the predicted  maximum of the
pion cloud contribution~\cite{SL,KY,SS}. Thus this experiment
is ideally suited to test these calculations.

The coincident $p(\vec{e},e^{\prime}\pi)$ cross section in
the one-photon-exchange-approximation can be written
as~\cite{obs}

\begin{eqnarray}
\label{eq:xsection}
\frac{d\sigma}{d\omega d\Omega_{\rm e} d\Omega^{\rm cm}_{\pi}}
    &=& \Gamma_{\rm v}~\sigma_{\rm h}(\theta,\phi),
\end{eqnarray}
\noindent where
\begin{eqnarray*}
	\sigma_{\rm h}(\theta,\phi) &=&
	\sigma_{\rm T} + \varepsilon \sigma_{\rm L} +
	\sqrt{2\varepsilon(1+\varepsilon)} \sigma_{\rm TL} \cos\phi\\ \nonumber
 	&\quad& + \varepsilon \sigma_{\rm TT}\cos 2\phi 
 + h p_{\rm e}\sqrt{2\varepsilon(1-\varepsilon)}\sigma_{\rm
TL^{\prime}} \sin\phi, \quad 
\end{eqnarray*}

 \noindent
$\Gamma_{\rm v}$ is the virtual photon flux, $h = \pm 1$ is
the electron helicity, $p_{\rm e}$ is the magnitude of the
longitudinal electron polarization,
$\varepsilon$ is the virtual photon  polarization,
$\theta $ and $\phi$  are the pion CM polar and azimuthal
angles relative to the momentum transfer $\vec{q}$, and
$\sigma_{\rm L}$, $\sigma_{\rm T}$, $\sigma_{\rm TL}$, and $\sigma_{\rm
TT}$ are the longitudinal, transverse, transverse-longitudinal, and
transverse-transverse interference cross sections,
respectively~\cite{obs}.

The TL and the TL$^{\prime}$ (transverse-longitudinal) cross sections
are the real and imaginary parts of the same combination of interference
multipole amplitudes. Approximate expressions for these are:

\begin{eqnarray}
\label{eq:TL}
	\sigma_{\rm TL}(\theta) &=&
		- \sin\theta {\rm Re}[ A_{\rm TL} +B_{\rm TL}
\cos\theta],
	\\
	\sigma_{\rm TL'}(\theta) &=&
		\sin\theta {\rm Im}[ A_{\rm TL} +B_{\rm TL}
\cos\theta],
	\nonumber \\
	A_{\rm TL} &\approx& -\frac{q}{k}L_{0+}^{*}M_{1+}, \nonumber
\\
	B_{\rm TL} &\approx& -6\frac{q}{k}L_{1+}^{*}M_{1+}, \nonumber
\end{eqnarray}

\noindent
where the pion production multipole amplitudes
are denoted by $M_{l\pm }$, $E_{l\pm }$, and $L_{l\pm }$,
indicating their character (magnetic, electric, or
longitudinal), their total angular momentum ($J$=$l\pm
1/2$), $q$ and $k$ are the pion and photon center of mass momenta.  In the first two lines  of Eq.2
it has been assumed that the pions are produced in s and p waves only. In the next two lines an additional truncated multipole approximation is made, namely only terms which interfere with the dominant magnetic dipole amplitude $M_{1+}$ are kept. The exact formulas without this approximation can be found in~\cite{obs}. In model calculations~\cite{SL,KY,MAID,Az} 
this approximation is not made and significant deviations from 
the truncated multipole approximation occur. 

As has been previously demonstrated $\sigma_{\rm TL}$ is
sensitive to the magnitude of the longitudinal quadrupole
amplitude C2~\cite{Mertz}. Adding a measurement of
$\sigma_{\rm TL^{\prime}}$ to this provides a stringent test of the
background magnitudes and phases of the reaction calculations. It
should be pointed out that a determination of the background
amplitudes is an important part of the physics of the
$\gamma \pi$N system.

To precisely determine the resonant quadrupole amplitude in
the $\gamma^* N\rightarrow\Delta$ transition at low $Q^{2}$
and to address the issue of background contributions, a
program has been developed at the MIT-Bates Linear
Accelerator. For this purpose we have developed a special
out-of-plane magnetic spectrometer system (OOPS) in which
the spectrometers are deployed symmetrically about the
momentum transfer $\vec{q}$ ~\cite{OOPS}. We observed the
TL$^{\prime}$ cross section using a polarized electron beam
of 0.85\% duty factor at an energy of 820 MeV. A typical polarization
and an average current were 37\% and 6 ${\rm \mu A}$, respectively.
A liquid H$_{2}$ target was
used in a cylindrical cell of 1.6 cm diameter with a 4.3 $\mu$m-thick
Havar wall.  The scattered electrons were detected in the
One Hundred Inch Proton Spectrometer (OHIPS)
and the coincident protons in two out-of-plane
spectrometers deployed at a fixed laboratory angle relative
to $\vec{q}$ and with out-of-plane angles
$\phi = 225^{\circ},~315^{\circ}$.
The focal plane instrumentation of each spectrometer
consisted of three horizontal drift chambers for track
reconstruction and scintillators for triggering.
Detailed optics studies were done for each spectrometer, and
the detection efficiencies were measured as functions of all independent
reaction coordinates.  The total efficiency of the system was
calibrated by using elastic electron scattering data from the liquid
H$_{2}$
target. Boiling effects in the target were studied by varying the
beam current and they were negligible. 
The phase-space normalization of the cross section and
various corrections applied to the data, including radiative corrections,
were calculated with a Monte Carlo simulation.
The cross sections were obtained from the part of the phase space of
the two spectrometers which were matched in four dimensions 
($W,Q^{2},\theta,\phi$)~\cite{Kunz}, where $W$ is a central invariant mass.

The experiment was performed at $Q^{2}= 0.127$ (GeV/c)$^{2}$,
$W = 1232$ MeV and 1170 MeV.
The results and kinematic settings are presented in
Table 1 and in Figs. 1, 2, and 3. The
helicity asymmetry $A_{\rm TL'}(\theta,\phi)$ is
\begin{eqnarray}
\label{eq:Ah}
	A_{\rm TL'}(\theta,\phi)
    &=&
		\frac{ \sigma_{h = 1}(\theta,\phi)
		    -  \sigma_{h= -1}(\theta,\phi) }
		     { \sigma_{h = 1}(\theta,\phi)
	            +  \sigma_{h= -1}(\theta,\phi) },\\ \nonumber
    &=&
		\frac{ \sqrt{2\varepsilon(1-\varepsilon)}
			\sigma_{\rm TL^{\prime}}(\theta)\sin(\phi) }
		     { \sigma_{\rm unpol}(\theta,\phi) }
\end{eqnarray}
\noindent
where the quantities were defined in Eq.~\ref{eq:xsection}
and $\sigma_{\rm unpol}$ is the electron helicity independent part
(the first four terms) of $\sigma_{\rm h}$. To first
approximation this quantity can be extracted from the data
without  detailed Monte Carlo calculations of the phase
space acceptance of the apparatus, and  therefore has a
smaller error. There are only small corrections due to the finite
acceptances of the spectrometers. In the quoted results the
quantities have been referred to the central spectrometer
settings~\cite{Kunz}. The absolute values of the TL$^{\prime}$ cross
sections
$\sigma_{\rm TL^{\prime}}$ have also been extracted from the
data~\cite{Kunz} and are presented here.

We note that the sign of $A_{\rm TL^{\prime}}$ is negative. In the $\pi^{0}$
channel the recoil protons were detected with the protons
being emitted in the forward direction relative to
$\vec{q}$, the momentum transfer with out of plane angles
$\phi_{\rm pq} = 45^{\circ},~135^{\circ}$.
In Eq.~\ref{eq:xsection} the angles of the pion are involved,
and $\phi= 225^{\circ},~315^{\circ}$. Therefore a negative sign
for $A_{\rm TL^{\prime}}$ means that the sign of $\sigma_{\rm TL^{\prime}}$
is positive.

The experimental results are compared to calculations
~\cite{SL,KY,MAID,Az} in Figs. 1, 2, and 3. The most ambitious
calculations are the Sato-Lee model~\cite{SL}
which calculates all of the 
multipoles and $\pi-N$ scattering from dynamical equations.
It is in agreement with the photoproduction data (some of the model parameters were fit to these data).  The Sato-Lee model also agrees with our data for the unpolarized cross sections 
$\sigma_0=\sigma_{\rm T}+\epsilon \sigma_{\rm L}$ but unfortunately is in strong disagreement with our measurements of $\sigma_{\rm TL}$ and $\sigma_{\rm TL^{\prime}}$. 
The dispersion relations calculation~\cite{Az} agrees 
with some of our data but disagrees with our measurement of $\sigma_{0}$ at
$W$ = 1170 MeV and with our $\sigma_{\rm TL}$ measurements. 
On the other hand dispersion relation calculations 
provide good agreement with photo-pion production 
data~\cite{Mainz-dispersion}. 
The Mainz Unitary Model (MAID) is a flexible way to fit observed cross sections as a function of $Q^{2}$~\cite{MAID}. 
It incorporates Breit-Wigner resonant terms, Born
terms,  higher $N^{*}$ resonances, and is unitarized using empirical
$\pi-$N phase shifts. The parameters of the model have been previously fit to a range of 
data,  including our previous results~\cite{Mertz}, and are in reasonable agreement with our data~\cite{MAID} with the exception of $\sigma_{\rm TL}(\theta=
119^{\circ}, W = 1170$ MeV) which is in disagreement with all
calculations. 
The Dubna - Mainz - Taipei (DMT) model~\cite{KY} includes dynamics for the resonant channels 
and uses the background amplitudes of  the MAID model. This model is in
reasonable agreement with our data at resonance ($W$ = 1232 MeV) but not with $\sigma_{0}$ and $\sigma_{\rm TL}$ below resonance ($W$ = 1170 MeV).

The Sato-Lee and DMT dynamical models~\cite{SL,KY} predict that the pion 
cloud is the dominant contribution to the quadrupole amplitudes at low values of $Q^{2}$. This behavior is an expected consequence of the spontaneous breaking of chiral symmetry in QCD~\cite{AB}. Unfortunately these models are not in overall agreement with our
data. In contrast
the Sato-Lee model showed much better predictions of the recently reported
JLab Hall B result for the $p(\vec{e}, e' p)\pi^{0}$  reaction
in the $\Delta$ region from $Q^{2}$ from 0.4 to 1.8 (GeV/c)$^{2}$~\cite{Joo}.
This seems to indicate that the dominant meson cloud contribution, which
is predicted to be a maximum near our values of $Q^{2}$, is not
quantitatively correct.  

Recently a measurement of $A_{\rm TL^{\prime}}$ for the
$p(\vec{e}, e' p)\pi^{0}$  reaction in
the $\Delta$ region was performed at Mainz~\cite{MainzTL'}. The kinematics
include a range of $Q^{2}$ values from 0.17 to 0.26 (GeV/c)$^{2}$ and
backward $\theta$ angles. These data were only compared to the MAID model with which they disagreed.

It is of interest to compare the TL and TL$^{\prime}$ results presented here
with those of the recoil polarizations which are
proportional to the real and imaginary parts of interference
multipole amplitudes. For the $p(\vec{e},e' \vec{p})\pi^{0}$
channel the outgoing proton polarizations have been observed in parallel kinematics (the protons emitted along $\vec{q}$ or $\theta = 180^{\circ}$)
~\cite{warren,Mpolp}. For this case the observables in the truncated multipole approximation are:

\begin{eqnarray}
\label{eq:Pn}
\sigma_{0}\: p_{x}&\propto&Re[A_{TL}^{x}], \\ \nonumber
\sigma_{0}\:p_{y}&\propto& Im[B_{TL}^{y}], \\ \nonumber
\sigma_{0}\:p_{z}&\propto& Re[C_{TT}^{z}], \\ \nonumber
A_{TL}^{x}\simeq
B_{TL}^{y}&\approx&(4L_{1+}^{*} -L_{0+}^{*}+ L_{1-}^{*}) M_{1+}, \\ \nonumber
C_{TT}^{z}&\approx& \mid M_{1+}\mid^{2}+Re[(6E_{1+}^{*}-2E_{0+}^{*})M_{1+}],
\end{eqnarray}
\noindent
where $\sigma_{0}$ is the unpolarized cross section, $p_{x}$, $p_{y}$, and $p_{z}$ are defined in~\cite{MAID}, and the constants  of proportionality contain only kinematic factors(for the full expressions see~\cite{obs}). This shows both the similarity and detailed difference
between a measurement of TL and TL$^{\prime}$ and the recoil polarizations.  In the published papers~\cite{warren,Mpolp} the data were compared to the MAID model which is found not to be in good agreement with the data. At the present time we do not have sufficient data to pin down the multipoles that are responsible for this difference (a discussion of the data requirements for
model independent analyses is presented in~\cite{AB}).  On the other hand there is a possible experimental problem since the data do not agree with a model independent sum rule~\cite{tiator_sumrule}.

The empirical MAID calculation gives the best overall agreement with the
accurate data presented here but also with the overall set of the data
obtained by our collaboration.  The parameters of this model for the
values of the resonant multipoles are 
$|M_{1+}(I=3/2)|= (40.9 \pm0.3)10^{-3}/m_{\pi}$, 
$CMR= C2/M1= -6.5 \pm 0.3\%$, 
$EMR=E2/M1=-2.2 \pm 0.9\%$.  
The errors are experimental and were obtained by varying the magnitudes 
in MAID of the resonant amplitudes by one $\sigma$ in a $\chi^2$ fit 
to our data.
In a previous paper~\cite{Mertz} we showed that the dominant error was due to model uncertainties, which was estimated by taking the differences between the multipoles of the different models. We argue here that only one model is in reasonable agreement with our data, and so this approach can no longer be used. Therefore the model errors come from uncertainties in the non-resonant multipole amplitudes of the MAID calculation, which at the present time are not known. We plan to address this issue in a future publication. 

In conclusion, we have performed the first simultaneous measurement of both
the real and imaginary parts of the transverse-longitudinal interference (TL) cross section for the $p(\vec{e}, e' p)\pi^{0}$ reaction in
the $\Delta$ region at a low $Q^{2}$ where the meson cloud is predicted to be the leading cause of deformation. It is found that only the more empirical MAID model~\cite{MAID}
is in reasonable agreement with the accurate data obtained for
$\sigma_{0}= \sigma_{T} + \epsilon \sigma_{L}$ and $ \sigma_{TL^{'}}$
and also with our previous data~\cite{Mertz}. On the other hand, for
recoil polarization data obtained  at the same
$Q^{2}$~\cite{warren,Mpolp}, and for $\sigma_{TL^{'}}$ taken at  slightly
higher values of $Q^{2}$~\cite{MainzTL'}, there  are possible problems
with the MAID model (results for other models were not presented in
these publications). At the present time there are not sufficient data
to ascertain which multipoles might be responsible for this
situation. We are presently analyzing new data taken at $Q^{2} = 0.127$
(GeV/c)$^{2}$ which should provide a more accurate determination of
these values~\cite{Sparveris} in conjunction with the new generation, double
polarization experiments~\cite{Buuren}.

We would like to thank the Bates staff for making this experiment possible. 
This work is supported in part by the US Department of Energy and the 
National Science Foundation, Grant-in-Aid for International
Scientific Research by the Ministry of Education, Science, and Culture 
in Japan, and Deutsche Forschungsgemeinschaft, the Humboldt Foundation, 
and the ESOP.


\newpage

\begin{figure}[p]
\begin{center}
\epsfig{file=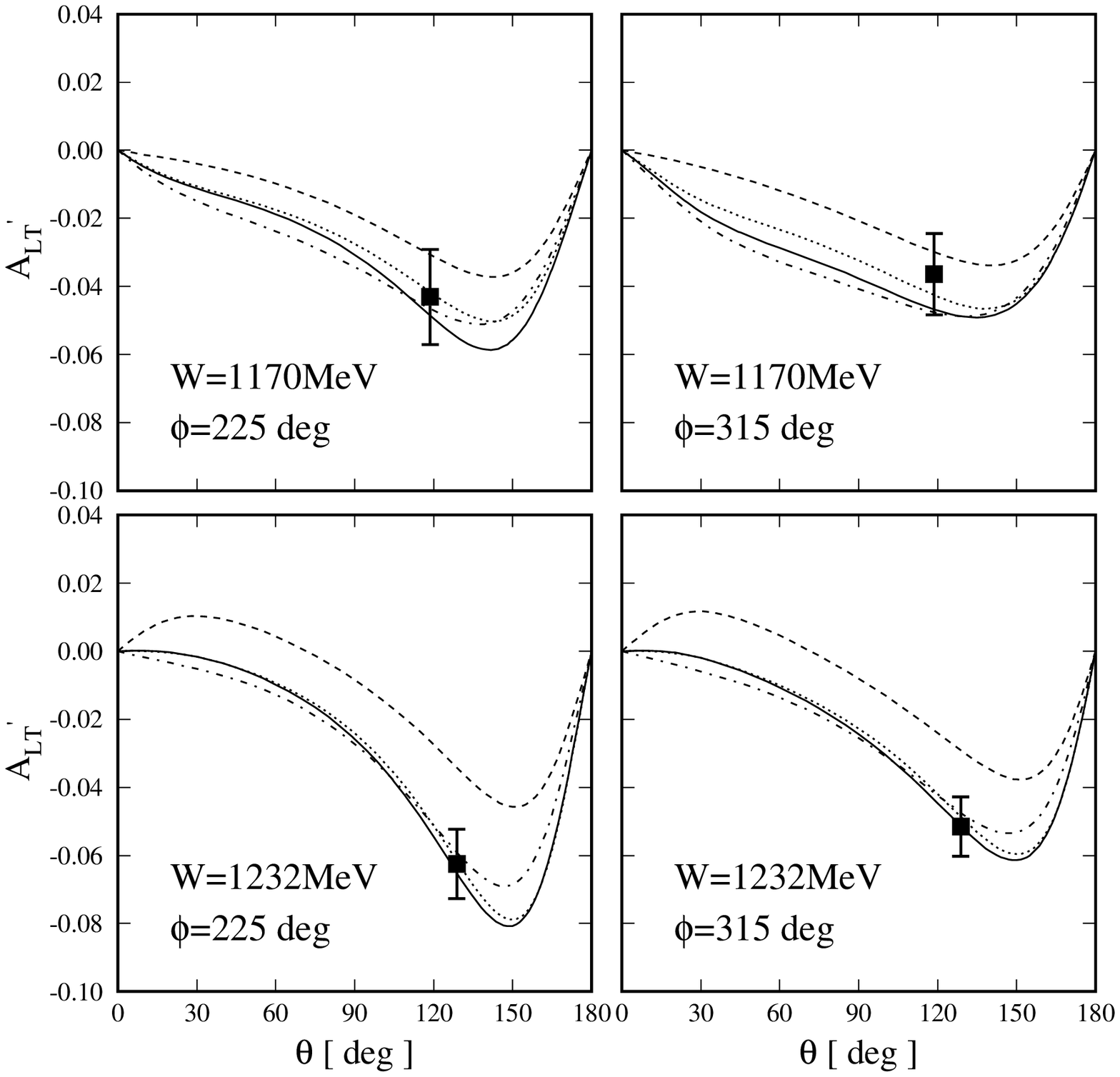,width=8cm}
\end{center}
\caption{The helicity asymmetry $A_{\rm TL'}$ for the
$p(\vec{e}, e' p)\pi^{0}$ reaction at $Q^{2} = 0.127$ (GeV/c)$^{2}$
plotted versus $\theta$ the CM angle between
the outgoing pion and the momentum transfer
$\vec{q}$. The curves are MAID~\cite{MAID} (solid), Sato-Lee~\cite{SL} (dashed), DMT~\cite{KY} (dotted), dispersion theory~\cite{Az} (dot-dashed).}
\end{figure}

\begin{center}
\begin{table}
\caption{Results of the present $p(\vec{e},e^{\prime}p)\pi^0$ experiment
at $Q^{2}=0.127$ (GeV/c)$^{2}$.}
\begin{tabular}{ccccc}
\hline
$W$ (MeV) & $\theta$ & $\sigma_{0}$ ($\mu$b/sr)
& $\sigma_{\rm TL}$ ($\mu$b/sr) & $\sigma_{\rm TL^{\prime}}$($\mu$b/sr)
\\
\hline
1170 & $119^{\rm \circ}$ & $17.31\pm 0.90$ & $0.91\pm 0.18$ &  $1.65\pm
0.55$ \\
1232 & $129^{\rm \circ}$ & $26.39\pm 0.47$ & $2.83\pm 0.20$ &  $3.11\pm
0.55$ \\
\hline
\end{tabular}
\label{tab:results}
\end{table}
\end{center}

\begin{figure}[p]
\begin{center}
\epsfig{file=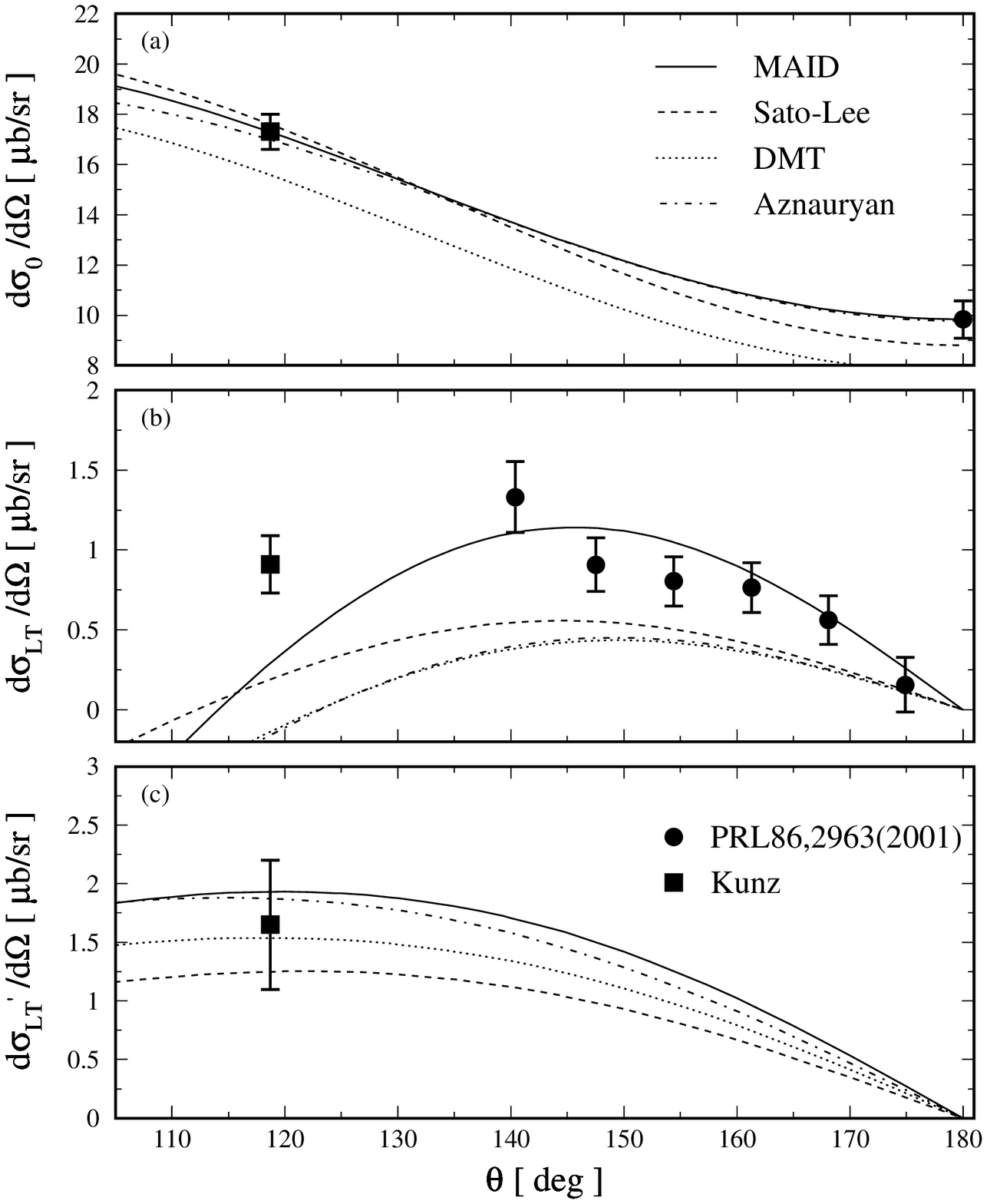,width=8cm}
\end{center}
\caption{
Cross sections for the $p(\vec{e}, e' p)\pi^{0}$ reaction for
$W = 1170$ MeV, $Q^{2} = 0.127$ (GeV/c)$^{2}$ plotted
versus $\theta$. Panel (a) is for $\sigma_{0} = \sigma_{\rm T} +
\epsilon \sigma_{\rm L}$. Panel (b) is for $\sigma_{\rm TL}$
and panel (c) is for $\sigma_{\rm TL^{\prime}}$.
 See the Fig. 1 captions for an explanation of the curves.
}
\end{figure}

\begin{figure}[p]
\begin{center}
\epsfig{file=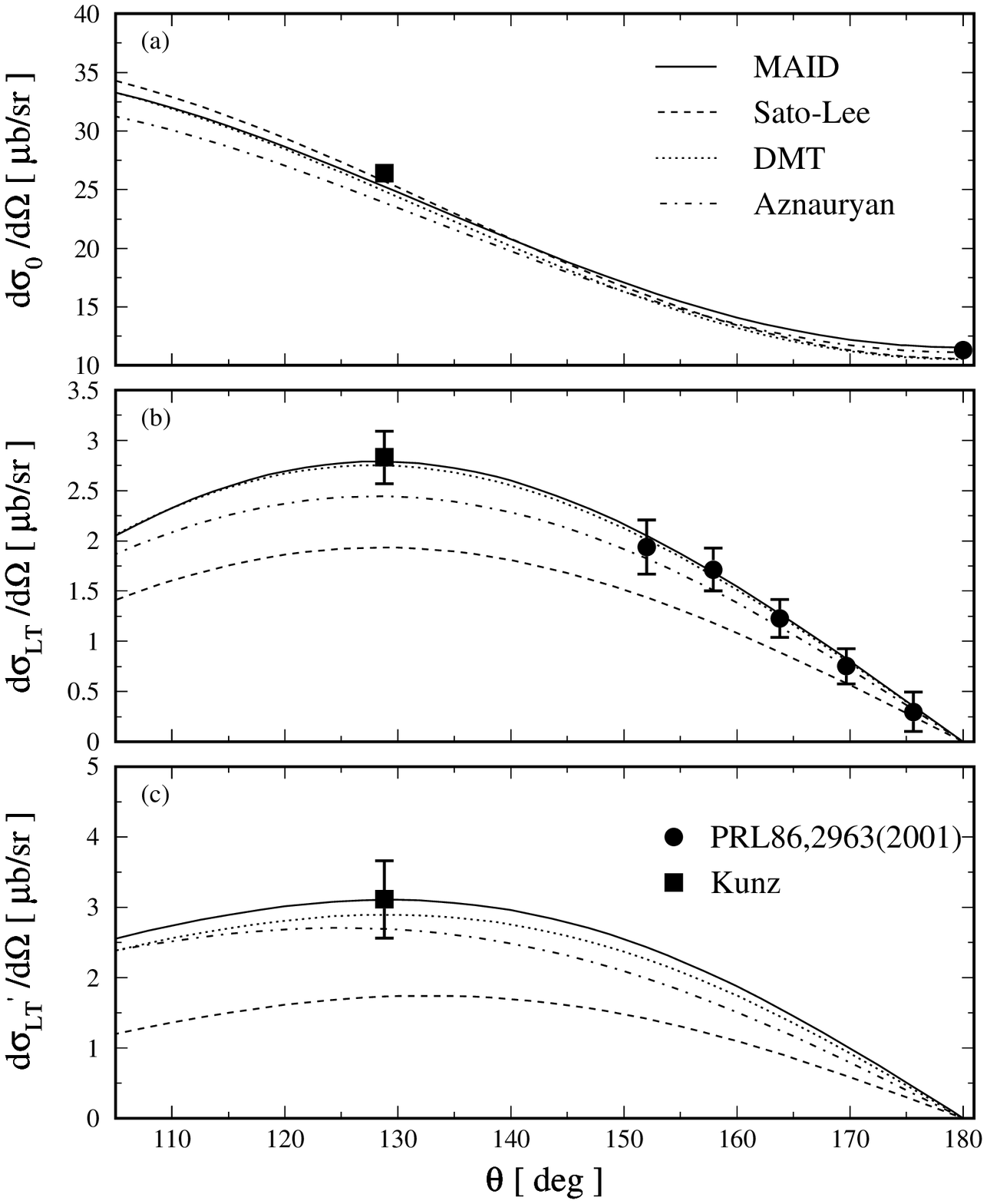,width=8cm}
\end{center}
\caption{
Cross sections for the $p(\vec{e}, e' p)\pi^{0}$ reaction for
$W = 1232$ MeV, $Q^{2} = 0.127$ (GeV/c)$^{2}$ plotted
versus $\theta$. Panel (a) is for $\sigma_{0} = \sigma_{\rm T} +
\epsilon \sigma_{\rm L}$. Panel (b) is for $\sigma_{\rm TL}$
and panel (c) is for $\sigma_{\rm TL^{\prime}}$.
See the Fig. 1 captions for an explanation of the curves.}
\end{figure}



\begin{thebibliography}{99}

\bibitem{Nstar}
	See e.g. NStar 2001, Proceedings of the Workshop on the 
	Physics of Excited Nucleons, D. Drechsel and L. Tiator editors, 
	World Scientific (2001).
\bibitem{Glashow} 
	S.L.~Glashow, Physica {\bf 96A}, 27 (1979).
\bibitem{LEGS} 
	G.~Blanpied {\it et~al.}, Phys. Rev. {\bf C64}, 025203 (2001).
\bibitem{Mainz} 
	R.~Beck {\it et~al.}, Phys. Rev. {\bf C61}, 35204 (2000).
\bibitem{Mertz}
	C.~Mertz {\it et~al.}, Phys. Rev. Lett. {\bf 86}, 2963 (2001).
\bibitem{CNP}
        C.N.~Papanicolas, International Conference on Quark Nuclear Physics, 
        Julich, Germany, June 9-14 (2002).
\bibitem{AB} A.M.Bernstein, Invited talk at Electron-Nucleus
	Scattering VII, June 23-28,2002, Elba, Italy, hep-ex/0212032.
\bibitem{CEBAF} Jefferson Laboratory experiment E91-011, S. Frullani, J. Kelly and A. Sarty spokesmen.
\bibitem{Vellidis} C.~Vellidis, Ph. D. 
thesis, University of Athens (2001), ISBN 960-8313-05-8.
\bibitem{OOPS} S.M.~Dolfini {\it et al.}, Nucl. Instrum. Methods.  {\bf A344}, 571 (1994), J.~Mandeville {\sl et al.}, Nucl. Instrum. Methods. {\bf A344}, 583 (1994), 
Z.-L.~Zhou {\it et al.}, Nucl. Instrum. Methods.  {\bf A487}, 365 (2002).
\bibitem{Isgur}  
	N.~Isgur, G.~Karl, and R.~Koniuk, Phys. Rev. {\bf D25}, 2394 (1982).
\bibitem{Goldstone} 
	J.~Goldstone, Nuovo Cimento {\bf 19}, 154 (1961). 
\bibitem{SL} 
	T.~Sato and T.-S.~H.~Lee, Phys. Rev. {\bf C63}, 055201 (2001).
\bibitem{KY}
	S.S.~Kamalov and S.N.~Yang, Phys. Rev. Lett. {\bf 83}, 4494 (1999).
\bibitem{SS} 
	M.~Fiolhais {\it et~al.}, Phys. Lett. {\bf B373}, 229 (1996).
\bibitem{obs}
	D.~Drechsel and L.~Tiator, J. Phys. G: Nucl. Part. Phys. {\bf 18},
	 449 (1992), A.S.~Raskin and T.W.~Donnelly, Ann. Phys. {\bf 191},  
	78 (1989).
\bibitem{MAID}
	D.~Drechsel {\it et~al.}, Nucl. Phys. {\bf A645}, 145 (1999) 
	and http://www.kph.uni-mainz.de/MAID/ 
\bibitem{Az} 
	I.G.~Aznauryan, Phys. Rev. {\bf D57}, 2727(1998) and
  	private communications.
\bibitem{Kunz} 
	C.~Kunz, Ph.D. Thesis, M.I.T.(2000), unpublished.
\bibitem{Mainz-dispersion} 
	O.~Hanstein, D.~Drechsel, and L.~Tiator
	Nucl. Phys. {\bf A632}, 561 (1998).
\bibitem{Joo} K.~Joo {\it et~al.}, Phys. Rev. Lett.{\bf 88}, 122001(2002).
\bibitem{MainzTL'} 
	P.~Bartsch {\it et al.}, Phys. Rev. Lett.{\bf 88},142001(2002).
\bibitem{warren} 
	G.A.~Warren {\it et~al.}, Phys. Rev. {\bf C58}, 3722 (1998).
\bibitem{Mpolp} 
	Th.~Pospischil {\it et al.}, Phys. Rev. Lett.{\bf 86}, 2959 (2001).
\bibitem{tiator_sumrule}
	H.~Schmieden and L.~Tiator, Eur.\ Phys.\ J.\  {\bf A8}, 15 (2000).
\bibitem{Sparveris} N.~Sparveris, Ph.D. Thesis, 
	University of Athens (2002), unpublished. 
\bibitem{Buuren}  L.~D.~van~Buuren {\it et al.}, 
	Phys. Rev. Lett.{\bf 89}, 012001 (2002),
	Proceedings of the Second Workshop, on
  Electromagnetic Physics with Internal Targets and the BLAST Detector, R.
  Alarcon and R.Milner editors, World Scientific, Singapore(1999).
\end{thebibliography}
\end{document}